# Exploring the experiences of undergraduate physics students taking a "Teaching Physics in School" module and the effects on their intentions to become a secondary physics teacher

**D Cottle**

School of Physics and Astronomy, University of Birmingham, Birmingham, UK

d.cottle@bham.ac.uk

**Abstract**. This article describes an undergraduate physics module at a university in England that places year 3 student physicists in secondary school classrooms for a semester. This is done as a way of introducing them to the occupation of secondary physics teaching using a realistic job preview approach. The module helps the undergraduate students develop their communication and professional skills and supports the physics learning of pupils in the schools where they are placed. The perceptions of the participating students toward secondary school physics teaching are then investigated. Two themes emerged from this research: The difficulty of explaining physics concepts to children and the need to make school physics enjoyable.

## 1. Introduction

Recruitment of new secondary school physics teachers in England in the recent past has been the lowest on record and is on a concerning downward trend. In 2020-21 only 22% of government target were recruited [1] and in the following academic year by September 2022 this figure was 15% [2]. This means that by this point, only 398 new physics teachers had been recruited against an increased target of 2610 (Teach First recruitment is excluded from this data). To put these numbers in context, in the same academic year in England a total of 5,620 students graduated from physics and astronomy degrees [3] making the target for physics teacher recruitment an unrealistic 46% of the total, but the actual number recruited a disappointing 7%. Similar shortages of physics and STEM teachers have been reported around the world [4] suggesting that it is important to fully understand ways of introducing secondary teaching as a career possibility for physics undergraduate students.

In England, physics teacher recruitment efforts have often focused on offering financial incentives for initial training, but the evidence indicates that these make only a limited and not a lasting impact [5,6,7]. The literature does suggest some more effective strategies on improving recruitment which are: Giving potential physics teachers the opportunity to see if the job is the 'right fit' for them through realistic job previews [5]. Demonstrating how their love for the physics can address social justice issues [8]. Offering support for them as new physics teachers [6]. Against this worrying backdrop, this article will explore a much needed and realistic approach to encourage physics undergraduates to consider a career as a physics teacher that connects them with the school physics education community, provides a realistic experience of teaching physics and supports what is known from literature about physics undergraduates' motivations to teach.

## 2. Realistic Job Preview

Realistic job preview attempts to support individuals in finding out whether a particular occupation will meet their own preferences and needs in employment. The intention of this being that if the right people are able to select the right occupation it will lead to long term job satisfaction for the individual and lower rates of job turnover for the employing organisation [10]. A realistic job preview differs from a more traditional approach to recruitment in that it aims to show job candidates both positive and negative aspects of a role in an attempt to ensure they have realistic expectations rather than relying on a potentially distorted overly optimistic presentation of reality that may lead to later disappointment [11]. Similar strategies in the US [12] have been shown to contribute to increasing the recruitment of new physics teachers. Situating the introduction secondary physics module in the undergraduate physics degree program is also significant as it has been found in Sweden that culture and assumptions of physics departments can influence students' choices to train to teach physics [13].

## 3. Context and Description of the Module

The study took place in a selective research-intensive university in England. In the School of Physics in each year of the undergraduate physics course there is a cohort of around 170 students. In the teacher education department of the School of Education there is a post-graduate physics specialist training program where the annual cohort varies between 4 and 12 students. Around a third of this number are recruited annually from the School of Physics. Strong links have been made between the School of Physics and the School of Education to facilitate the delivery of the teaching physics in school module with a joint academic post allowing a specialist physics education member of academic staff to teach the module.

### 3.1 Selection of Students

The teaching physics in schools module takes the form of a year 3, optional 10 credit module. This is one of many other available options, all others of which are traditional lecture-based courses culminating in written examinations (E.g. Condensed matter physics or chaos and dynamical systems). Information about it and the other modules is presented to students during the spring term of year 2 in a lecture format allowing them to meet the module lead and get further details. The module is unlike other options because main component of student learning happens on a weekly 3 hour per week school placement in a local secondary school physics classroom. Other differences are that there is an application process for students to be selected for the module: Assessment is by written assignment and verbal presentation rather than examination, and students must attend training on professional skills in the previous semester to prepare them to go to school in their new role. Students must also successfully obtain a mandatory security check called a DBS (UK government disclosure and barring service check) to regularly work with under 18's in schools. A follow up meeting for further information and questions is also provided so that students can understand the module fully before applying. There are two purposes of the need for selection: the number of places on the module is limited by the number of school placements available and the module is very different in its purpose, structure and assessment than other physics modules. Therefore it is important that students who are best suited for the type of learning and the opportunities provided by the module are able to take it. The application form asks prospective students to give reasons for their choice and to provide any details of other experience they have of volunteering in schools or working with young people. Prospective students are selected by their demonstrable desire to investigate physics teaching as a possible future career. Suitable candidates are then interviewed by the module lead. This takes the form of asking them to prepare a short 10 minute 'lesson' on a school physics topic which they present, followed by questions. One of the module aims is to develop the communication skills of students so judgements are not made on the quality only but also on the ability of the students to relate to young people and their and their suitability to engage in a realistic job preview placement in a local secondary school.

### 3.2 Training

After selection, training is provided to the successful students. This takes the form of an in-person 3 hour workshop with activities based around understanding the context of local schools, examples of physics learning situated in the school curriculum, attention to the change of role from pupil to role-model and teacher. An important part of this is an understanding of safeguarding and the processes in place in local schools. Students are given contact details of a physics teacher in their school who will act as their point of contact and mentor. They are asked to contact them by email to arrange a meeting to have a preliminary preparatory discussion to find out information about the school. Regular weekly communication about the module occurs online from the module tutor and there is another in-person workshop to promote community and reflect on practice mid-way through the semester.

### 3.3 Assessment

Assessment for the module takes the form of eight weekly reflection sheets submitted to the module tutor electronically. These are a way of keeping in regular touch with the students while they are on placement and to engage the students in dialogue about their experiences and learning. The reflections allow for personalized support to be offered by the module tutor. A small amount of academic credit is awarded for these submissions to encourage participation. At the end of the module students are asked to present in front of their peers for 15 minutes to explain their own learning from their time in school and demonstrate their developing verbal communication and teaching skills (%). There is also 2000-word written assignment that students must also write on a topic mutually agreed between themselves, the module tutor and their school mentor. This assignment needs to focus on an aspect of physics education and must involve some aspect of data gathering from the placement school. For example, by observing lessons or asking pupils or teachers questions. The choice of topic is chosen by the student based on their own interests. Example titles of such assignments are provided to students and recently have included: Gender stereotypes in the physics classroom, An investigation into the preferences of GCSE Physics pupils for the creation of a resource to introduce the topic of electric circuits, How can we introduce recent discoveries in physics in a relevant and engaging way to enthuse future physicists? A final component of assessment involves asking the school mentors to provide formative comments on various aspects of professional practice and learning to the students.

## 4. Secondary School Placements

Supportive school placements within a reasonable travelling distance of the university are essential for the delivery of this module. Strong links between the School of Physics and the School of Education are again leveraged to find suitable schools and often these are the same schools and mentors that offer places for trainee secondary physics teachers. This overlap is significant in reinforcing the purpose of the placement and providing a supportive and realistic experience for the students similar to that they might experience while formally training to be a teacher. School mentors are provided with a handbook explaining the requirements of the module and are contacted annually to ascertain the capacity of their school to offer places in the coming year. Local schools and colleges normally offer a mix of places for one or two students. The administrative burden for school mentors is deliberately kept as low as possible to support reducing the workload of serving teachers. It consists of meeting with the student before the placement begins to ensure they are aware of all relevant school policies and arrangements for travel, professional dress and so on and to arrange a convenient time for the placement to occur on a weekly basis. They also agree a timetable with the student of classes they will attend and support on their placement. It is suggested that this is on a Wednesday afternoon, but this is not always convenient and students and schools are able to negotiate a time between themselves that meets the minimum of 3 hours spent in school. This allows some flexibility for students to work with different age pupils or fit in with their own study commitments rather than needing to accept whatever is on the school timetable on Wednesday afternoons. This is a physics module, so the schools are asked to provide the students with physics lessons rather than general science as a priority. Sometimes however a compromise has to be made based on time constraints and general science classes are used. The activity and experience of each student on placement is not specified exactly as flexibility accommodates the variety of opportunities available in different schools and the particular learning needs of the students. Typical experiences include: lesson observation, lesson support in a teaching assistant type role, student led demonstrations of practical activities, student support with extra-curricular science clubs, delivering careers talks, supporting 6th form pupils with university applications, teaching short segments of lessons with the support of teachers. School and students are both aware that the student teacher is not a member of staff and is also not a trainee teacher – they are there in a voluntary and supportive capacity and should not be put in any position of exclusive responsibility for pupil safety, wellbeing or learning. During the placement some students also receive a monitoring visit from the module tutor – this takes the form of an informal visit mainly to check on wellbeing and reinforce strong links with the schools rather than assessment.

## 5. Methodology of Study

A short online survey of attitudes to teaching secondary school physics (Figure. 1) was administered to students taking the module who consented to participate in the study (n=5). This was done twice; before they started and after the end of the module to investigate any changes in attitudes. Students were also

**Teaching Physics in School Survey Questions**

On a scale of 1-10 (1 – low, 10 – high) please indicate:

1. Your impression of a secondary physics secondary school teacher as a career valuable to society.
2. Your view of a secondary physics secondary school teacher as a career where your physics knowledge will be useful.
3. Your impression of a secondary physics secondary school teacher as a career you could be successful at.
4. Your impression of a secondary physics secondary school teacher as a career you can see yourself fitting in.
5. Your impression of a secondary physics secondary school teacher as a career where you can make a difference.
6. How likely it is you will train to be a secondary school teacher at the end of your undergraduate physics course.
7. How likely it is you will train to be a secondary school teacher at some point during your career.

Free text questions:

8. Please explain in a few sentences why you chose to take the Teaching Physics in School module.
9. Please explain in a few sentences your impressions of what it is like to be a secondary school physics teacher.

**Figure 1.** Attitudinal survey questions given to undergraduate physics students.

asked if they would participate after the end of the module in a semi-structured interview conducted via video call about their experiences on the module and how it had affected their intentions to become a physics teacher. Two students volunteered and the interviews were transcribed and thematically analysed [14]. Communication with students about their participation in the study was done by an academic colleague not connected with the module. This was an attempt to ensure students were not influenced to participate by a desire to gain credit in module assessment, although the two interviews were conducted by the module lead after module was complete. The themes discussed in the interview were:

- Describe the types of things a physics teacher does and are these similar or different to what you thought before the module began.
- What do you think are the biggest challenges in physics teaching?
- Describe a critical incident on your school placement – a key moment that caused a change in your thinking.
- Describe any of your skills that you think have been developed through participating in this module with examples.
- How likely are you to train to be a physics teacher in the future?

## 6. Findings

Due to the small number of participants in this study the data from the survey questions were combined in figure 2 to show that there was no significant change overall in their attitudes toward physics teaching before and after completing the module. Breaking down individual questions showed a similar pattern with the exception of agreement with the statement that physics teaching is 'a career valuable to society'. Applying a one-tailed Matt-Whitney U-test due to the small numbers of participants, non-parametric data and assumption that participation in the module would increase positive attitudes to physics teaching, the median value increased from 8 to 9 (Mann-Whitney $U$=3.5 $n_1$=$n_2$=5 $p<0.05$). This is interpreted as a possible small increase. Responses to the open questions are summarized in table 1.

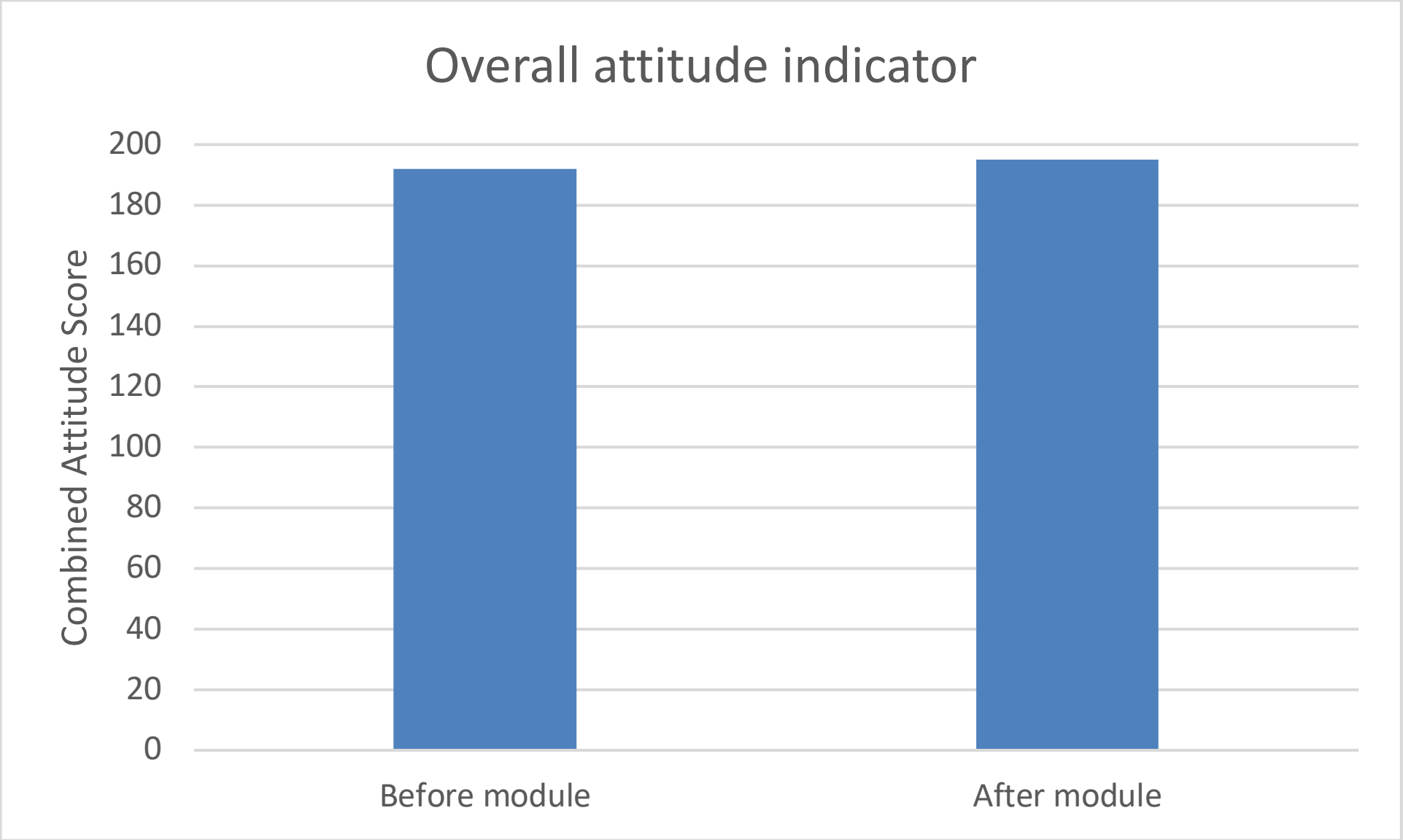

**Figure 2**. Overall attitude indicator from survey administered before and after students had completed the physics teaching in schools module.

These show consistent feelings that the module helps the students find out if teaching is the right career for them, but there are also interesting changes in responses before and after the school placement. There are no references to a sense of fun or enjoyment in the responses after the end of the placement and there is the addition of comments related to recognition of the personal skills developed, communication and confidence being named. Another new response after the module is that this module was different to others. Responses to the question about what it is like to be a physics teacher so a trend toward more responses acknowledging the difficulties of the job with words used like 'thankless', 'draining' and 'requires patience' in addition to 'hard' and 'stressful' that were anticipated before.

**Table 1.** Survey Free Text Question Responses.

| Question | Pre-placement | Post placement |
|---|---|---|
| Why did you choose the module? | Fun (2)<br>Personal satisfaction<br>To find out if teaching is for me (4)<br>Classroom experience (2)<br>Use physics knowledge | To find out if teaching is for me (4)<br>Module different to other options (2)<br>Improve communication skills<br>Improve confidence |
| What is it like to be a physics teacher? | Busy<br>Interesting<br>Varied<br>Hard<br>Stressful<br>Rewarding (2)<br>Adapting physics knowledge for children | Hard (3)<br>Stressful / draining (2)<br>Rewarding (2)<br>Requires patience<br>Thankless<br>Enables children to learn physics |

In terms of the key question, how likely is it you will train to be a secondary school teacher at some point during your career there was no significant change of view although the free text responses add context. One participant changed their mind about teaching, deciding it was now not for them. They described the experience of physics teachers as “hard work, emotionally draining, and thankless” and so decided that “I'm not cut out for it”. Two participants however used the experience to confirm their choice of career. “I think it is a profession where I will get a lot of job satisfaction. It is a job which I am confident I can do well and make an impact into other's lives. Also, there is a national shortage of science teachers so I feel like I would be contributing towards a greater standard of science education by becoming a science teacher.” It “felt worthwhile when you had positive experiences with students and made you feel like you are actually having an impact on someone’s learning.” One participant was still undecided and described a family members experience “My mum was a teacher and had to leave the profession because it was (un) sustainable for her in terms of her stress levels and enjoyment of work despite knowing the impact and positive parts of her job”.

Data from the semi-structured follow up interviews with two participants was transcribed, coded and them thematically analysed. Table 2 shows the process whereby two overarching themes were identified by the researcher as ‘Difficulty of explaining physics to children’ and ‘Encouraging children to enjoy physics’. The relationship between these overlapping themes and the contributing comments are shown in Figure 3.

**Table 2.** Thematic analysis of interview data.

| Question | Codes | | Themes |
|---|---|---|---|
| | Interview 1 | Interview 2 | |
| What does a physics teacher do? | Uses physics knowledge<br>Understands children<br>Helps children<br>Mentors children<br>Cares for children<br>Promotes physics as a subject<br>Explains physics concepts | Uses physics knowledge<br>Explains physics concepts<br>Inspiring children<br>Promoting physics as a subject<br>Positive impact on children | Difficulty of explaining physics to children<br><br>Encouraging children to enjoy physics |
| What are biggest challenges in physics teaching? | Getting children interested in physics<br>Not knowing what will happen everyday<br>Pupil behaviour management<br>Administrative issues in school | Gender bias<br>Ensuring conceptual understanding of physics<br>Making physics enjoyable | |
| Critical incident | Challenge of explaining physics concepts to children | Helping students who struggle to understand physics<br>Pupil behaviour management | |
| Skills developed | Verbal communication<br>Written communication<br>Pupil behaviour management | Confidence<br>Verbal communication<br>Explaining physics concepts<br>Professionalism | |
| Intention to teach physics in future | Gave time to consider teaching as a career<br>Stressful<br>Increased desire to teach | Inadequate pay<br>Not put off teaching | |
| Other comments | | Importance of modelling a struggle with physics<br>Empathy<br>Stressful | |

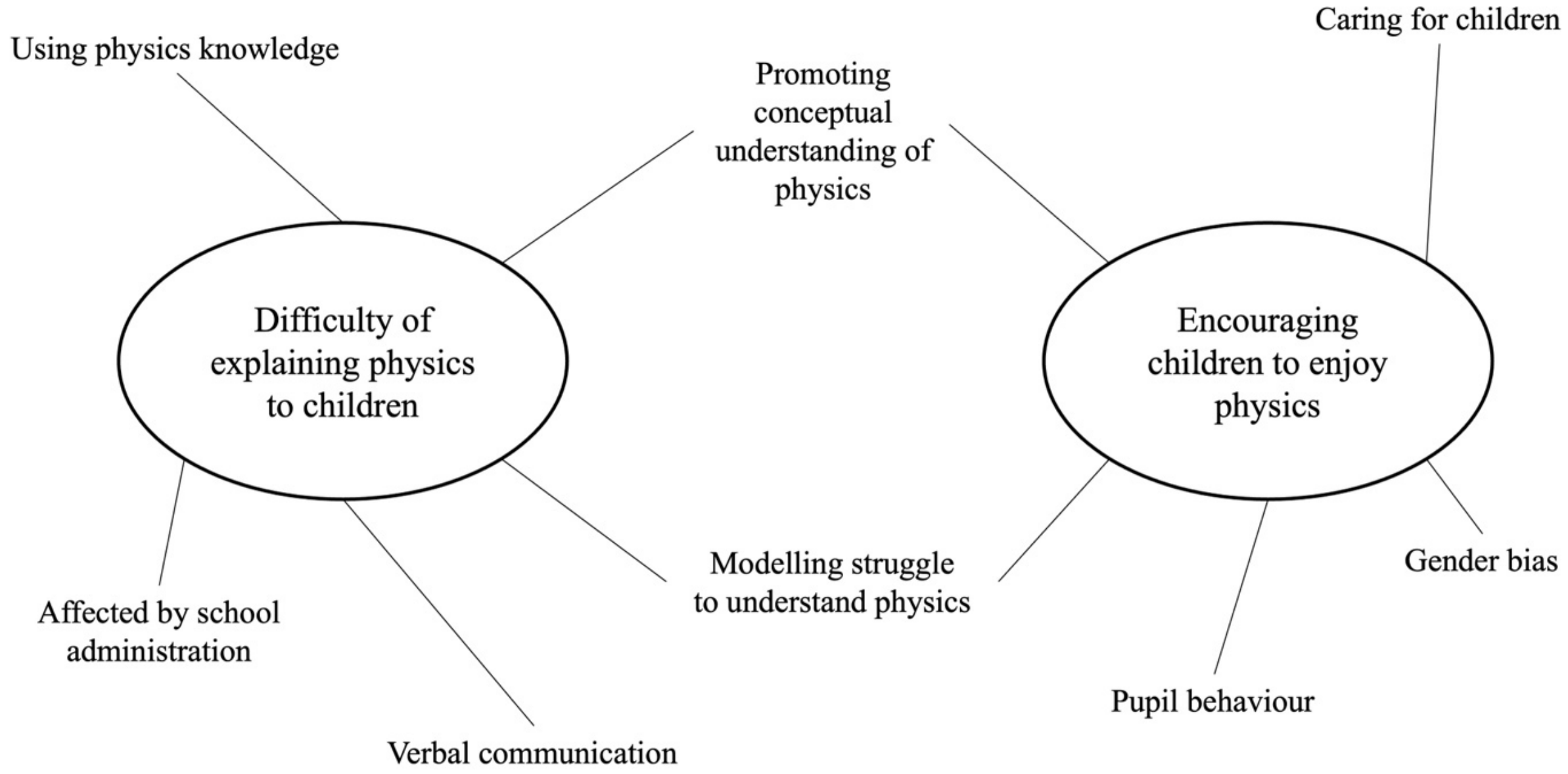

**Figure 3.** Summary of thematic analysis.

## 7. Discussion

According to literature discussed above, the framework of the realistic job preview approach sets out to prevent new entrants to a profession having unrealistically high expectations that later contribute to feelings of dissatisfaction [10]. The teaching physics in schools module described above succeeds as a realistic job preview in that students expectations of hard work and the stressful aspects of teaching physics were confirmed. The module does not showcase exiting and positive parts of the teaching experience in an attempt to entice undergraduate physics students into the teaching career despite a temptation to do so arising from the current shortage. Instead attitudes to teaching remain the same for participating students before and after the module with hints of improvements only in terms of how much they can see that physics teachers are needed in society. Areas of difficulty observed by students on their placements around negative pupil behaviour, onerous school administrative demands, inadequate pay and workplace stress are familiar but importantly are also realistic parts of the experience of teaching physics. Comments from the interviews coalesce into two similar themes; the need to explain physics to children and the difficulty of doing so. Perhaps the question in the minds of these undergraduate physics students with a prior positive disposition toward teaching is: Do the benefits of physics teaching outweigh the challenges?

Another interesting theme emerging is the desire to make physics enjoyable for children. Recent literature has questioned the extent to which school physics in England is 'for' a broad range of pupils of different gender, ethnic or socio-economic backgrounds [15] due to perceptions of physics as difficult, elitist or even intrinsically masculine. The students on this module showed a natural awareness of these issues, raising them unprompted. One student discussed the importance of modelling a 'struggle' with physics. That is, showing that physics is not effortless but that engaging with problems, perhaps working collaboratively on solutions is a pleasurable and an intrinsic part of the enjoyable experience of physics. Perhaps there are hints of changes in attitude here of future physics teachers that may tentatively promise some improvement in widening access to physics for a broader range of pupils?

There is therefore some evidence from this limited study that the experiences undergraduate physics students have on this module are helping them to think carefully about their career choice – both toward and away from teaching. It is hoped that the effect of the module may be to support retention in the profession by those who do choose a physics teaching career because they will better informed about the realities they will encounter. The physics teaching in school's module requires a significant amount of commitment from the students in terms of time and energy – much more so on a consistent week by week basis than one of the other 10 credit optional modules available to them. This may mean that the group of students who had opted for the Teaching Physics in Schools module in year 3 of their undergraduate physics degree were likely to be predisposed to positive attitudes toward teaching. A comparison of attitudes to physics teaching with a group of students not choosing the teaching physics in schools module would therefore be useful.